\documentclass[lettersize,journal]{IEEEtran}
\usepackage{amsmath,amsfonts}
\usepackage{algorithmic}
\usepackage{algorithm}
\usepackage{array}
\usepackage{textcomp}
\usepackage{stfloats}
\usepackage{url}
\usepackage{verbatim}
\usepackage{graphicx}
\usepackage{cite}
\usepackage[utf8]{inputenc}
\usepackage{mathtools}
\usepackage{booktabs}
\usepackage{multirow}
\usepackage{amsmath}

\usepackage{caption}
\usepackage{subcaption}
\usepackage{printlen}
\usepackage{setspace} % for "\setstretch" macro

\usepackage{tabularx}
\usepackage{bm}
\usepackage{array}
\usepackage{epstopdf}
\usepackage{float}
\usepackage{stfloats}
\usepackage{xcolor}
\usepackage{adjustbox}
\usepackage{csquotes}
\usepackage{multicol}
\usepackage{multirow}
\usepackage{cuted}
\usepackage{amsthm}
\allowdisplaybreaks % allow page breaks, if needed

\usepackage{ifthen}
\usepackage{amssymb}% http://ctan.org/pkg/amssymb
\usepackage{pifont}% http://ctan.org/pkg/pifont
\begin{document}

\title{Connected Cruise and Traffic Control \\ for Pairs of Connected Automated Vehicles}

\author{Sicong~Guo,~G{\'{a}}bor~Orosz,~and~Tamas~G.~Molnar%xxx,~\IEEEmembership{Member,~IEEE,}
        % <-this % stops a space
%\thanks{This paper was produced by the IEEE Publication Technology Group. They are in Piscataway, NJ.}% <-this % stops a space
%\thanks{Manuscript received April 19, 2021; revised August 16, 2021.}
\thanks{This research was supported by the University of Michigan’s Center for Connected and Automated Transportation through 
the US DOT grant 69A3551747105.}
\thanks{S. Guo and G. Orosz are with the Department of Mechanical Engineering, University of Michigan, Ann Arbor, MI 48109, USA (e-mail: stevengu@umich.edu, orosz@umich.edu).}
\thanks{G. Orosz is also with the Department of Civil and Environmental Engineering, University of Michigan, Ann Arbor, MI 48109, USA.}
\thanks{T. G. Molnar is with the Department of Mechanical and Civil Engineering, California Institute of Technology, Pasadena, CA 91125, USA (e-mail: tmolnar@caltech.edu).}
}

% The paper headers
%\markboth{Journal of \LaTeX\ Class Files,~Vol.~14, No.~8, August~2021}%
%{Shell \MakeLowercase{\textit{et al.}}: A Sample Article Using IEEEtran.cls for IEEE Journals}

%\IEEEpubid{0000--0000/00\$00.00~\copyright~2021 IEEE}
% Remember, if you use this you must call \IEEEpubidadjcol in the second
% column for its text to clear the IEEEpubid mark.

\maketitle

\begin{abstract}
This paper considers mixed traffic consisting of connected automated vehicles equipped with vehicle-to-everything (V2X) connectivity and human-driven vehicles.
A control strategy is proposed for communicating pairs of connected automated vehicles, where the two vehicles regulate their longitudinal motion by responding to each other, and, at the same time, stabilize the human-driven traffic between them.
Stability analysis is conducted to find stabilizing controllers, and simulations are used to show the efficacy of the proposed approach.
The impact of the penetration of connectivity and automation on the string stability of traffic is quantified.
It is shown that, even with moderate penetration, connected automated vehicle pairs executing the proposed controllers achieve significant benefits compared to when these vehicles are disconnected and controlled independently.
\end{abstract}


%%%%%%%%%%%%%%%%%%%%%%%%%%%%%%%%%%%%%%%%%%%%%%%%%%%%%%%%%
\section{Introduction}
%%%%%%%%%%%%%%%%%%%%%%%%%%%%%%%%%%%%%%%%%%%%%%%%%%%%%%%%%

%%%%%%%%%%%%%%%%%%%%%%%%%%%%%%%%%%%%%%%%%%%%%%%
\begin{figure*}[b]
	\centering
    \includegraphics[scale=1]{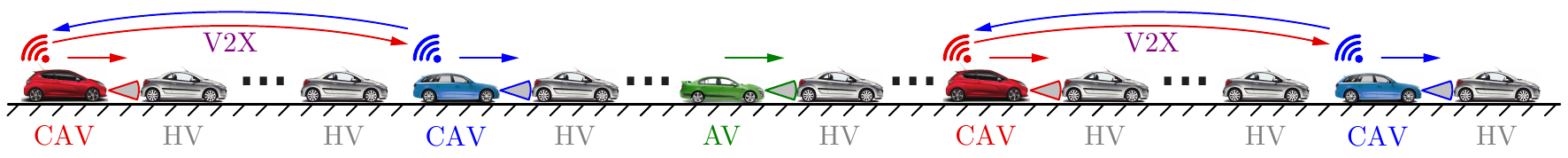}
    \caption{Mixed traffic consisting of human-driven vehicles (HVs), automated vehicles (AVs), and pairs of connected automated vehicles (CAVs).
    The pairs of CAVs enclose vehicle packets that they can directly regulate with the proposed controllers in order to stabilize traffic.
    }
    \label{fig:setup}
\end{figure*}
%%%%%%%%%%%%%%%%%%%%%%%%%%%%%%%%%%%%%%%%%%%%%%%

Vehicle automation continues to gain ground thanks to its potential for driving efficiency, safety and comfort.
Recently, the field of longitudinal control for automated vehicles (AVs) has seen a surge in research activity.
Several works have addressed {\em adaptive cruise control (ACC)} from aspects like safety~\cite{Niletal2016,AmesXuGriTab2017}, string stability~\cite{Gunter2021arecommercially}, personalization of driving behavior~\cite{Work2021}, and compensation of response delays~\cite{Bekiaris-Liberis2018}, while ACC systems have become widely available to the public.

Apart from a single vehicle, traffic control~\cite{Siri2021, Ferrara2018} may also benefit from automation through the positive impact of AVs on large-scale traffic~\cite{Cicic2018, Piacentini2019}.
AVs can act as mobile actuators to improve traffic smoothness~\cite{Cui2017, Yu2018, Zheng2020, Giammarino2021, Hayat2022, Bayen2022, Stern2022} that was demonstrated by experiments~\cite{Stern2018}.
This helps to mitigate traffic congestion and thereby reduce pollutant emissions and noise.
Moreover, it has been shown that controlling platoons of AVs may further improve the flow of traffic~\cite{Karafyllis2021, Feng2021, Cicic2022}.

As such, many works have focused on vehicle platoons that yield potential for cooperation.
Specifically, platoons of connected automated vehicles (CAVs) equipped with vehicle-to-everything (V2X) connectivity may exchange information and execute {\em cooperative adaptive cruise control (CACC)}~\cite{johansson2017cooperative, Bertoni2017, McAuliffe2018, wang2018review_CACC, wouw2019MPC}.
While CACC has significant positive impact on traffic~\cite{Shladover2012, Silgu2021}, its disadvantage is that it requires full penetration of connectivity and automation within an entire platoon.

Thus, interest has arisen in studying mixed traffic with lean penetration of CAVs amongst connected human-driven vehicles (CHVs).
On one hand, a CAV may connect to CHVs ahead of it, and obtain information beyond its line of sight for use in control.
This strategy, called {\em connected cruise control (CCC)}~\cite{Orosz2016, Zhang2016}, has outperformed ACC in experiments~\cite{Ge2018}.
On the other hand, the CAV may also connect to CHVs behind it, and use the information from connectivity to control and stabilize the following traffic.
This approach was used as {\em connected traffic control (CTC)}  in~\cite{Molnar2020cdc, Molnar2022ch}, leading cruise control (LCC) in~\cite{wang2020lcc, wang2022deeplcc}, and considerate model predictive control in~\cite{Ard2022}.
With these strategies, connectivity can bring great benefits for both the CAV and the following vehicles, ultimately leading to safer, smoother, string stable traffic.

The benefits of connectivity have been shown clearly by the literature above.
Still, a sufficient penetration of connectivity is required for these benefits~\cite{Avedisov2022}.
Yet, connectivity is voluntary: the owners of human-driven vehicles (HVs) may decide not to invest in V2X devices and stay disconnected.
At the same time, the cost of establishing communication is marginal compared to that of automation, hence it is more likely that AVs will be upgraded to CAVs than that HVs become CHVs.

Therefore, instead of investigating CAV platoons or connectivity between CAVs and CHVs, this paper focuses on mixed traffic where {\em pairs of CAVs} get connected while traveling amongst HVs.
Connectivity allows the two CAVs to cooperate and respond to each other in a mutually beneficial manner, while controlling and stabilizing the traffic enclosed by them.

%%%%%%%%%%%%%%%%%%%%%%%%%%%%%%%%%%%%%%%%%%%%%%%%%%%%%%%%%
\subsection{Concept, Contributions and Benefits}
%%%%%%%%%%%%%%%%%%%%%%%%%%%%%%%%%%%%%%%%%%%%%%%%%%%%%%%%%

In this paper, we consider the scenario shown in Fig.~\ref{fig:setup}, in which a traffic fleet executes car-following on a single lane of a straight road.
The traffic consists of human-driven vehicles (HVs) and connected automated vehicles (CAVs) equipped with vehicle-to-everything (V2X) connectivity.
The CAVs that are outside the communication range of other CAVs act as automated vehicles (AVs) without connectivity (green).
The CAVs that are within each other's communication range  form pairs and respond to each other (blue and red).

Specifically, we focus on controlling the CAV pair in Fig.~\ref{fig:setup} that encloses human-driven traffic.
We assume lean penetration of connectivity and automation, i.e., multiple HVs between the CAVs.
Our contributions are summarized as follows.
\begin{itemize}
\item {\em Connected cruise and traffic control} is proposed in which a pair of CAVs regulates its longitudinal motion while stabilizing the traffic between them.
\item Stability analysis is conducted to find stabilizing controllers, by accounting for the response delays of vehicles.
\item Simulations are performed for a single CAV pair and large-scale traffic including multiple CAV pairs.
\item
The effects of CAV penetration on the string stability of traffic and the associated benefits of connectivity are quantified via stability charts and simulations.
\end{itemize}

\begin{figure}
    \centering
    \includegraphics[scale=1]{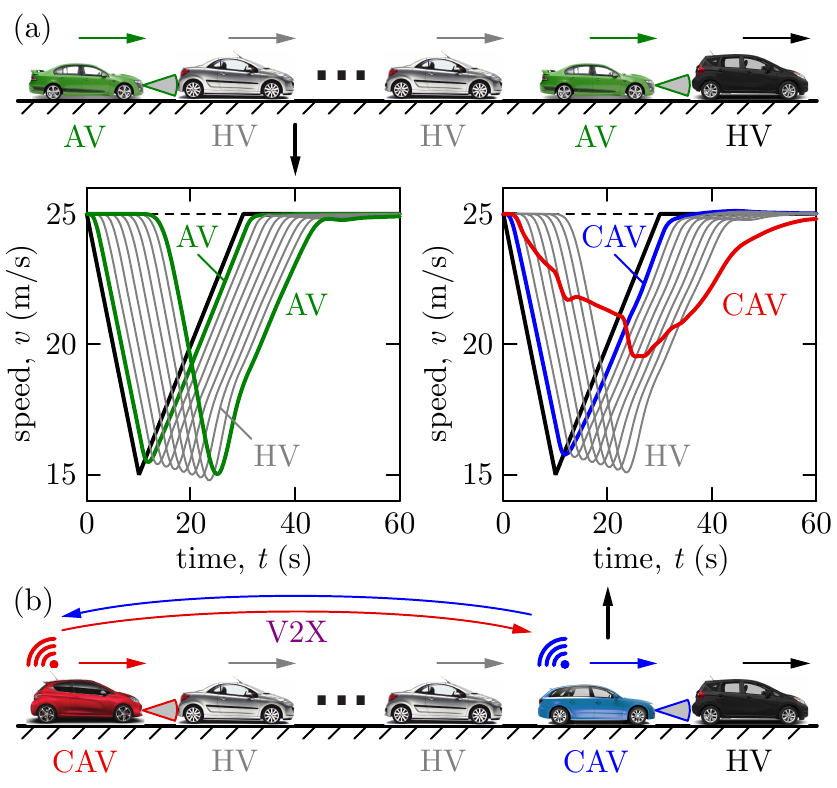}
    \caption{
    Dynamics of mixed traffic without and with connectivity.
    (a) Human-driven vehicles (HVs) and automated vehicles (AVs) executing adaptive cruise control.
    With string unstable HVs, the AVs fail to mitigate the onset of a congestion: the tail vehicle reduces its speed as much as the lead vehicle.
    (b) The proposed pair of connected automated vehicles (CAVs) executing connected cruise and traffic control.
    The CAVs successfully smoothen traffic: the tail vehicle reduces its speed much less than the lead.}
    \label{fig:benefits}
\end{figure}

To highlight the relevance of these contributions, we show that connectivity and the proposed control strategy for pairs of CAVs yield benefits compared to scenarios without connectivity.
These benefits are illustrated by an example in Fig.~\ref{fig:benefits}, where two traffic fleets without and with connectivity are compared via numerical simulations (with details given later).

%%%%%%%%%%%%%%%%%%%%%%%%%%%%%%%%%%%%%%%%%%%%%%%
%%%%%%%%%%%%%%%%%%%%%%%%%%%%%%%%%%%%%%%%%%%%%%%

Fig.~\ref{fig:benefits}(a) shows a heterogeneous chain of vehicles without connectivity, including a lead vehicle, an AV, 5 subsequent HVs, and another AV.
The lead vehicle brakes, accelerates and cruises at constant speed, while the subsequent vehicles respond.
The HVs exhibit {\em string unstable} behavior~\cite{Feng2019}, where they overreact to speed perturbations and reduce their speeds more than the vehicle ahead of them.
This undesired behavior may lead to traffic congestion, unless mitigated by others.
As opposed, the AVs behave string stable and reduce their speeds less than the vehicle preceding them.
Still, due to the relatively small number of AVs, the overall behavior is undesired: the tail vehicle reduces its speed as much as the lead.

Fig.~\ref{fig:benefits}(b) depicts the corresponding setup with connectivity, where a CAV pair responds to each other using our proposed controller.
Despite the string unstable human driving, the CAV pair successfully mitigates the onset of a congestion: the tail vehicle reduces its speed much less than the lead vehicle.
This {\em head-to-tail string stable}~\cite{Zhang2016} behavior is beneficial for traffic smoothness, travel times, and fuel consumption.

The details leading to these results are discussed as follows.
Section~\ref{sec:control} describes the proposed controllers, and the dynamical models of CAVs and HVs.
Section~\ref{sec:stability} discusses stability analysis.
Section~\ref{sec:casestudy} presents the results using stability charts and simulations, and quantifies the effect of CAV penetration.
Section~\ref{sec:conclusions} closes with conclusions.

%%%%%%%%%%%%%%%%%%%%%%%%%%%%%%%%%%%%%%%%%%%%%%%%%%%%%%%%%
\section{Control Design for Pairs of Connected Automated Vehicles}\label{sec:control}
%%%%%%%%%%%%%%%%%%%%%%%%%%%%%%%%%%%%%%%%%%%%%%%%%%%%%%%%%

In this section, we propose longitudinal controllers for pairs of connected automated vehicles (CAVs) traveling in mixed traffic, and to this end, we model the dynamics of CAVs and human-driven vehicles (HVs).

In particular, we focus on the vehicle packet highlighted in Fig.~\ref{fig:packet}(a), that travels on a single lane of a straight road.
The packet includes a pair of CAVs (called head and tail CAV, in blue and red) and $N$ number of HVs (gray). The packet travels behind a lead vehicle (labelled as HV, although it could be any vehicle type; see black).
We number the vehicles with indices increasing in the direction of motion, starting from the tail CAV with index 0.
We denote the headway of vehicle $i$ by $h_{i}$, and its velocity by $v_{i}$, ${i \in \{0, \ldots, N+2\}}$.

%%%%%%%%%%%%%%%%%%%%%%%%%%%%%%%%%%%%%%%%%%%%%%%%%%%%%%%%%
\subsection{Dynamics and Control of Connected Automated Vehicles}
%%%%%%%%%%%%%%%%%%%%%%%%%%%%%%%%%%%%%%%%%%%%%%%%%%%%%%%%%

\begin{figure}
    \centering
    \includegraphics[scale=1]{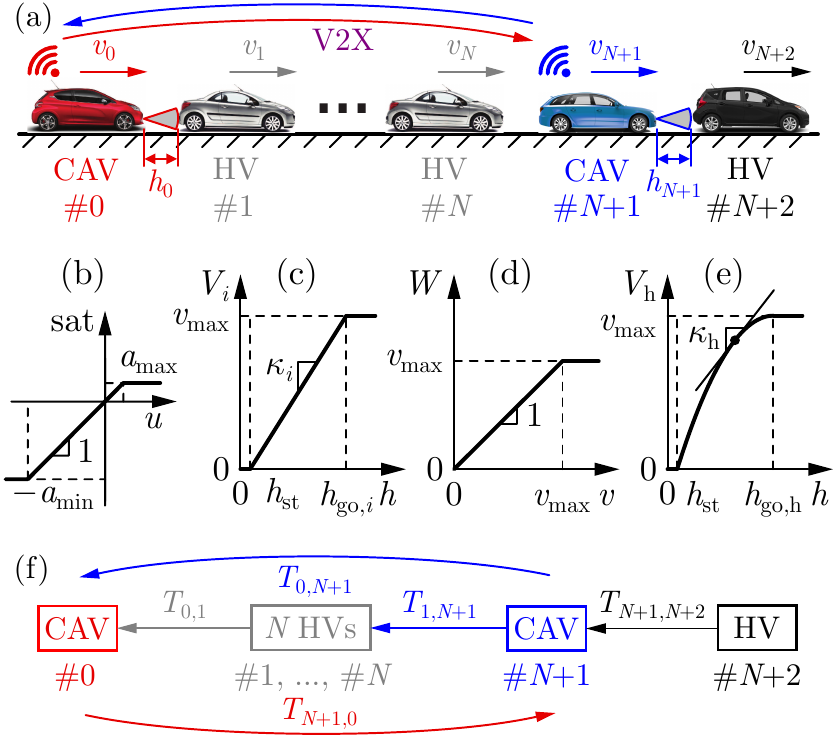}
    \caption{
    (a) A vehicle packet that consists of a pair of CAVs and HVs between them. The CAVs seek to stabilize this packet.
    (b)-(e) Nonlinearities in their longitudinal car-following dynamics.
    (f) Block diagram with the link transfer functions describing the responses of each vehicle.
    }
    \label{fig:packet}
\end{figure}

We capture the dynamics of CAVs by delayed double integrator models with saturation:
\begin{equation} \label{eq:CAV_model}
\begin{split}
\dot{h}_{0}(t)&=v_{1}(t)-v_{0}(t), \\
\dot{v}_{0}(t)&=\mathrm{sat}\big(u_{0}(t-\sigma_{0})\big), \\
\dot{h}_{N+1}(t)&=v_{N+2}(t)-v_{N+1}(t), \\
\dot{v}_{N+1}(t)&=\mathrm{sat}\big(u_{N+1}(t-\sigma_{N+1})\big), \\
\end{split}
\end{equation}
in which $u_{0}$ and $u_{N+1}$ are the desired accelerations of the tail and head CAV, respectively, that are considered as control inputs.
We assume that each CAV realizes the desired acceleration by low-level controllers, unless it is above the acceleration limit $a_{\max}$ or below the braking limit $-a_{\min}$.
This is captured by the saturation function:
\begin{equation}
    \mathrm{sat}(u)=\min{\{\max\{-a_{\min},u\},a_{\max}\}},
\label{eq:sat}
\end{equation}
shown in Fig.~\ref{fig:packet}(b).
Furthermore, we incorporate time delays $\sigma_{0}$ and $\sigma_{N+1}$ into the model to account for actuation, communication and feedback delays.
For simplicity, the dynamics of each CAV, including the parameters $a_{\max}$, $a_{\min}$ and ${\sigma_{0}=\sigma_{N+1}=\sigma}$, are assumed to be the same.

%%%%%%%%%%%%%%%%%%%%%%%%%%%%%%%%%%%%%%%%%%%%%%%
%%%%%%%%%%%%%%%%%%%%%%%%%%%%%%%%%%%%%%%%%%%%%%%

The focus of this paper is a longitudinal control strategy for the pair of CAVs that also allows stabilizing the traffic between them.
The control law is established based on~\cite{Zhang2016}, where {\em connected cruise control (CCC)} was proposed for CAVs to respond to connected (but not necessarily automated) vehicles ahead of them, and on~\cite{Molnar2022ch}, where {\em connected traffic control (CTC)} were introduced for CAVs to respond to connected vehicles behind them.
Our present work proposes a {\em connected cruise and traffic controller} that integrates CCC and CTC for pairs of CAVs, including the responses of both the tail CAV to the head CAV and vice versa.
This ultimately achieves benefits beyond those of controlling a single CAV.

The proposed controller involves the responses of the tail and head CAVs, respectively, to:
(i) the headways $h_{0}$ and $h_{N+1}$ ahead of them;
(ii) the velocities $v_{1}$ and $v_{N+2}$ of the vehicles preceding them; and
(iii) the velocities $v_{N+1}$ and $v_{0}$ of each other.
Specifically, the following control law is proposed:
\begin{equation} \label{eq:controller}
\begin{split}
u_{0}&=\alpha_{0}\big(V_{0}(h_{0})-v_{0}\big)+\beta_{0}\big(W(v_{1})-v_{0}\big)\\
&\quad+\beta_{0,N\!+\!1}(W(v_{N\!+\!1})-v_{0}), \\
u_{N\!+\!1}&=\alpha_{N\!+\!1}\big(V_{N\!+\!1}(h_{N\!+\!1})\!-\!v_{N\!+\!1}\big)+\beta_{N\!+\!1}\big(W(v_{N\!+\!2})\!-\!v_{N\!+\!1}\big)\\
&\quad+\beta_{N\!+\!1,0}\big(W(v_{0})-v_{N\!+\!1}\big), %+\beta_{(N+1),N}(W(v_{N}(t))-v_{N+1}(t))
\end{split}
\end{equation}
where $\alpha_{0}$, $\alpha_{N+1}$, $\beta_{0}$, $\beta_{N+1}$, $\beta_{0,N+1}$ and $\beta_{N+1,0}$ are control gains to be designed.
This corresponds to CCC for the tail CAV~\cite{Zhang2016} and CTC for the head CAV~\cite{Molnar2022ch}.
These controllers can be deployed on the CAVs in a decentralized fashion, while the gains can be jointly designed to leverage cooperation.

The first terms on the right-hand sides of~(\ref{eq:controller}) allow the CAVs to respond to the headways $h_{0}$ and $h_{N+1}$ by using the range policies $V_{0}$ and $V_{N+1}$ that prescribe a desired velocity based on the headway. 
We define these range policies by:
\begin{equation}\label{eq:range_policy_CAV}
    V_{i}(h)=
    \begin{cases}
    0 & h\leq h_{\mathrm{st}}, 
    \\
    v_{\max} \frac{h-h_{\mathrm{st}}}{h_{\mathrm{go},i}-h_{\mathrm{st}}}
    & h_{\mathrm{st}} < h < h_{\mathrm{go},i}, 
    \\
    v_{\max} & h \geq h_{\mathrm{go},i}, 
    \end{cases} 
\end{equation}
${i \in \{0, N+1\}}$; see Fig.~\ref{fig:packet}(c).
These command the CAVs to: stop if their headways are below the standstill headway $h_{\mathrm{st}}$; increase their speeds linearly for larger headways;
and travel at the speed limit $v_{\max}$ if their headways exceed the free-flow headways $h_{\mathrm{go},i}$.
Note that $h_{\mathrm{go}}$ can be designed to be different for the two CAVs, and smaller $h_{\mathrm{go}}$ yields more aggressive driving.
As such, the range policies $V_{i}$ of the CAVs are part of the control design.
We use the piecewise linear choice~(\ref{eq:range_policy_CAV}) because it has been tested in experiments extensively~\cite{Ge2018}, while other nonlinear range policy choices could also work.

The second terms on the right-hand sides of~(\ref{eq:controller}) involve response to the velocities $v_{1}$ and $v_{N+2}$ of the vehicles preceding the CAVs, or the speed limit, as given by the speed policy:
\begin{equation}
    W(v) = \min\{v,v_{\max}\};
\label{eq:W}
\end{equation}
cf.~Fig.~\ref{fig:packet}(d).
Finally, the third terms on the right-hand sides of~(\ref{eq:controller}) involve the response of the CAVs to each other, allowing them to coordinate their motion for traffic stabilization.
Note that setting ${\beta_{0,N+1} = 0}$ or ${\beta_{N+1,0} = 0}$ would eliminate these terms, yielding adaptive cruise control for the tail or head CAV, respectively, that could be implemented without connectivity.

%%%%%%%%%%%%%%%%%%%%%%%%%%%%%%%%%%%%%%%%%%%%%%%%%%%%%%%%%
\subsection{Human Driver Model}
%%%%%%%%%%%%%%%%%%%%%%%%%%%%%%%%%%%%%%%%%%%%%%%%%%%%%%%%%

Now we present a selected car-following model that captures the behavior of human-driven vehicles (HVs).
While the proposed controller~(\ref{eq:controller}) does not rely on a human driver model, this model will be used in our numerical case studies.

Similar to~(\ref{eq:CAV_model}), we model the dynamics of each HV as:
\begin{equation}\label{eq:HV_model}
    \begin{split}
        \dot{h}_{i}(t)&=v_{i+1}(t)-v_{i}(t), \\
        \dot{v}_{i}(t)&=\mathrm{sat}\big(u_{i}(t-\tau_{i})\big),
        \quad \forall i \in \{1, \ldots, N\},
    \end{split}
\end{equation}
where $u_{i}$ is the acceleration of vehicle $i$ commanded by the human driver, and
$\tau_{i}$ is the time delay that includes the actuation delay of the vehicle and the driver reaction time.

For simplicity of exposition, we consider HVs with identical driving behaviors, including the delay ${\tau_{i}=\tau}$ and the driver model that captures the commanded acceleration $u_i$.
Specifically, we use the {\em optimal velocity model}~\cite{Bando1998}:
\begin{equation}\label{eq:OVM}
    u_{i}=\alpha_{\rm h}\big(V_{\rm h}(h_{i})-v_{i}\big)+\beta_{\rm h}(v_{i+1}-v_{i}),
\end{equation}
where $\alpha_{\rm h}$ and $\beta_{\rm h}$ are driver parameters characterizing the response to the headway and the velocity of the preceding vehicle.
Similar to~(\ref{eq:controller}), the response to the headway is through the range policy $V_{\rm h}$, defined as:
\begin{equation}\label{eq:V_h}
\begin{split}
    V_{\rm h}(h)&=
    \begin{cases}
    0 & h\leq h_{\mathrm{st}}, 
    \\
    v_{\max}\frac{(2h_{\mathrm{go,h}}-h_{\mathrm{st}}-h)(h-h_{\mathrm{st}})}{(h_{\mathrm{go,h}}-h_{\mathrm{st}})^{2}} & h_{\mathrm{st}} \!<\! h \!<\! h_{\mathrm{go,h}}, 
    \\
    v_{\max} & h \geq h_{\mathrm{go,h}}.
    \end{cases}
\end{split}
\end{equation}
Based on the experimental results in~\cite{Avedisov2018}, this range policy is considered to be piecewise quadratic
as illustrated in Fig.~\ref{fig:packet}(e).

The results throughout this paper, including Fig.~\ref{fig:benefits}, are calculated using~(\ref{eq:CAV_model})-(\ref{eq:V_h}) and the parameters listed in Table~\ref{tab:parameters} (unless stated otherwise). For non-connected AVs, the special case $\beta_{0,N+1} = \beta_{N+1,0} = 0$ of~(\ref{eq:controller}) is used as ACC. 

%%%%%%%%%%%%%%%%%%%%%%%%%%%%%%%%%%%%%%%%%%%%%%%%%%%%%%%%%
\setlength{\tabcolsep}{6pt}
\begin{table}
\caption {Parameters of the numerical case study} \label{tab:parameters}
\begin{center}
	\small
	\begin{tabularx}{\columnwidth}{ccccc}
	    \toprule
	    Vehicle & Variable & Symbol & Value & Unit\\
    	\midrule
        \multirow{4}{*}{all}
    	& braking limit & $a_{\min}$ & $7$ & $\mathrm{m/s^{2}}$  
        \\ 
		& acceleration limit & $a_{\max}$ & $3$ & $\mathrm{m/s^{2}}$ 
        \\
    	& speed limit &$v_{\max}$ & $30$ & $\mathrm{m/s}$ 
        \\
		& standstill headway &$h_{\mathrm{st}}$ & $10$ & $\mathrm{m}$ 
        \\ 
		\midrule
    	CAVs & delay & $\sigma$& $0.6$ & $\mathrm{s}$  
        \\ 
		\midrule
        \multirow{5}{*}{tail CAV}
		& free flow headway &$h_{\mathrm{go},0}$ & $60$ & $\mathrm{m}$ 
        \\
		& range policy gradient & $\kappa_{0}$ & $0.6$ & $\mathrm{1/s}$ 
        \\ 
		& headway response gain &$\alpha_{0}$ & $0.4$ & $\mathrm{1/s}$ 
        \\ 
		  & speed response gain &$\beta_{0}$ & $0.5$ & $\mathrm{1/s}$ 
        \\ 
		  & speed response gain &$\beta_{0,N+1}$ & $0.8$ & $\mathrm{1/s}$ 
        \\
		\midrule
        \multirow{5}{*}{head CAV}
		  & free flow headway &$h_{\mathrm{go},N+1}$ & $60$ & $\mathrm{m}$ 
        \\
		& range policy gradient & $\kappa_{N+1}$ & $0.6$ & $\mathrm{1/s}$ 
        \\
		& headway response gain &$\alpha_{N+1}$ & $0.4$ & $\mathrm{1/s}$ 
        \\
		  & speed response gain &$\beta_{N+1}$ & $0.5$ & $\mathrm{1/s}$ 
        \\ 
		  & speed response gain &$\beta_{N+1,0}$ & $0.1$ & $\mathrm{1/s}$ 
        \\
		\midrule
        \multirow{5}{*}{HVs}
    	& delay & $\tau$& $0.8$ & $\mathrm{s}$  
        \\
		  & free flow headway &$h_{\mathrm{go,h}}$ & $60$ & $\mathrm{m}$ 
        \\ 
		  & range policy gradient &$\kappa_{\rm h}$ & $0.7$ & $\mathrm{1/s}$ 
        \\ 
		& headway response gain &$\alpha_{\rm h}$ & $0.1$ & $\mathrm{1/s}$ 
        \\ 
		& speed response gain &$\beta_{\rm h}$ & $0.6$ & $\mathrm{1/s}$ 
        \\
		\bottomrule
	\end{tabularx}
	\normalsize
\end{center}
\end{table}
%%%%%%%%%%%%%%%%%%%%%%%%%%%%%%%%%%%%%%%%%%%%%%%

%%%%%%%%%%%%%%%%%%%%%%%%%%%%%%%%%%%%%%%%%%%%%%%%%%%%%%%%%
\section{Stability Analysis}\label{sec:stability}
%%%%%%%%%%%%%%%%%%%%%%%%%%%%%%%%%%%%%%%%%%%%%%%%%%%%%%%%%

In this section, we study the dynamics of the vehicle packet shown in Fig.~\ref{fig:packet}(a).
We formalize stability conditions, analyze the head-to-tail string stability phenomenon in Fig.~\ref{fig:benefits}, and design controller parameters that achieve stable traffic.
The analysis is performed in Laplace domain after linearization, thus it yields local stability results w.r.t.~small velocity and headway perturbations around an equilibrium.
We will demonstrate the global nonlinear behavior afterwards via simulations.

%%%%%%%%%%%%%%%%%%%%%%%%%%%%%%%%%%%%%%%%%%%%%%%%%%%%%%%%%
\subsection{Linearized Dynamics}
%%%%%%%%%%%%%%%%%%%%%%%%%%%%%%%%%%%%%%%%%%%%%%%%%%%%%%%%%

We first linearize the dynamics and transform them to Laplace domain.
Linearization is done around the equilibrium: 
\begin{equation}
    v_{i}(t)\equiv v^{*},\quad h_{i}(t)\equiv h_{i}^{*},
    \quad \forall i \in \{0, \ldots, N+1\},
\end{equation}
where all vehicles drive with uniform equilibrium speed $v^{*}$, while keeping equilibrium headways $h_{i}^{*}$ that may be different for the individual vehicles, as given by:
$v^{*}=V_{0}(h_{0}^{*})=V_{N+1}(h_{N+1}^{*})=V_{\rm h}(h_{\rm h}^{*})$,
with ${h_{i}^{*} = h_{\rm h}^{*}}$ for ${i \in \{1, \ldots, N\}}$.

To construct the linearized dynamics, we consider perturbations around the equilibrium in the form: 
\begin{equation}
    v_{i}(t) = v^{*}+\Tilde{v}_{i}(t),\quad h_{i}(t) = h_{i}^{*}+\Tilde{h}_{i}(t),
\end{equation}
and collect these perturbations into the state vector $\mathbf{x}_{i}$:
\begin{equation}
    \mathbf{x}_{i}(t)=\begin{bmatrix}
 \Tilde{h}_{i}(t) \\ \Tilde{v}_{i}(t)
\end{bmatrix},
\end{equation}
from which the speed fluctuations can be obtained by:
\begin{equation}
    \Tilde{v}_{i}(t)=\mathbf{c}\mathbf{x}_{i}(t),\quad \mathbf{c} = \begin{bmatrix}
0 & 1 \\
\end{bmatrix}.
\end{equation}

We derive and analyze the linearized dynamics under the assumption that ${-a_{\min} < u_i < a_{\max}}$ and ${0 < v_i < v_{\max}}$, i.e., where nonlinearities in~(\ref{eq:sat}),~(\ref{eq:W}),~(\ref{eq:range_policy_CAV}) and~(\ref{eq:V_h}) are differentiable.
The corresponding linearized model reads:
\begin{equation}
\begin{split}
    \dot{\mathbf{x}}_{0}(t)&=\mathbf{a}\mathbf{x}_{0}(t)+\mathbf{a}_{0}\mathbf{x}_{0}(t-\sigma)
    \\
    &+\mathbf{b}\Tilde{v}_{1}(t)+\mathbf{b}_{0} \Tilde{v}_{1}(t-\sigma)+\mathbf{b}_{0,N+1}\Tilde{v}_{N+1}(t-\sigma), 
    \\
    \dot{\mathbf{x}}_{i}(t)&=\mathbf{a}\mathbf{x}_{i}(t)+\mathbf{a}_{\rm h}\mathbf{x}_{i}(t-\tau)
    \\
    &+\mathbf{b}\Tilde{v}_{i+1}(t)+\mathbf{b}_{\rm h}\Tilde{v}_{i+1}(t-\tau),
    \qquad \forall i \in \{1, \ldots, N\}, 
    \\
    \dot{\mathbf{x}}_{N\!+\!1}(t)&=\mathbf{a}\mathbf{x}_{N\!+\!1}(t)+\mathbf{a}_{N\!+\!1}\mathbf{x}_{N\!+\!1}(t-\sigma)
    \\
    &+\mathbf{b}\Tilde{v}_{N\!+\!2}(t) +\mathbf{b}_{N\!+1\!}\Tilde{v}_{N\!+\!2}(t-\sigma)
    +\mathbf{b}_{N\!+\!1,0}\Tilde{v}_{0}(t-\sigma),
    \end{split}
\label{eq:linearized_dynamics}
\end{equation}
with coefficient matrices listed in~(\ref{eq:linmat}) in the Appendix.
These matrices contain
${\kappa_{0}=\frac{{\rm d}V_{0}}{{\rm d}h}(h_{0}^{*})}$,
${\kappa_{N+1}=\frac{{\rm d}V_{N+1}}{{\rm d}h}(h_{N+1}^{*})}$ and
${\kappa_{\rm h}=\frac{{\rm d}V_{\rm h}}{{\rm d}h}(h_{\rm h}^{*})}$, that are the gradients of the range policies:
\begin{equation}\label{eq:kappas}
    \kappa_0 = \kappa_{N+1} = \frac{v_{\max}}{h_{\mathrm{go},i}-h_{\mathrm{st}}}, 
    \quad
    \kappa_{\rm h} = \frac{2v_{\max}(h_{\mathrm{go,h}}-h_{\rm h}^{*})}{(h_{\mathrm{go,h}}-h_{\mathrm{st}})^{2}},
\end{equation}
cf.~(\ref{eq:range_policy_CAV}),~(\ref{eq:V_h}) and Fig.~\ref{fig:packet}(c)-(e).
Note that according to \eqref{eq:range_policy_CAV} and \eqref{eq:kappas}
 the gradients $\kappa_{0}$ and $\kappa_{N+1}$ are associated with the equilibrium headways as
\begin{equation}\label{eq:kappas_vs_h}
h_{0}^{*}=\frac{v^{*}}{\kappa_{0}} + h_{\mathrm{st}}, \quad
h_{N+1}^{*}=\frac{v^{*}}{\kappa_{N+1}} + h_{\mathrm{st}}.
\end{equation}
Hence, larger range policy gradient means more aggressive driving with smaller equilibrium headway and more compact traffic.

We analyze the linearized dynamics~(\ref{eq:linearized_dynamics}) in Laplace domain by formulating {\em link transfer functions}~\cite{Zhang2016} associated with the responses of each vehicle.
With the link transfer function $T_{i,j}$, we   relate the speed perturbations (denoted by $\Tilde{V}$ in Laplace domain) of vehicles $i$ and $j$, as follows:
\begin{equation}
\begin{split}
    \Tilde{V}_{0}(s)&=T_{0,1}(s)\Tilde{V}_{1}(s)+T_{0,N+1}(s)\Tilde{V}_{N+1}(s),\\
    \Tilde{V}_{1}(s)&=\prod_{i=1}^{N}T_{i,i\!+\!1}(s)\Tilde{V}_{N\!+\!1}(s)\eqqcolon T_{1,N\!+\!1}(s)\Tilde{V}_{N\!+\!1}(s),\\
    \Tilde{V}_{N+1}(s)&=T_{N+1,0}(s)\Tilde{V}_{0}(s)+T_{N+1,N+2}(s)\Tilde{V}_{N+2}(s);
\end{split}
\label{eq:responses}
\end{equation}
see the block diagram in Fig.~\ref{fig:packet}(f).
Assuming zero initial conditions, the link transfer functions are obtained from~(\ref{eq:linearized_dynamics}):
\begin{align}\label{eq:sec3a-link-tf}
\begin{split}
    T_{0,1}(s)&= \mathbf{c}(s\mathbf{I}-\mathbf{a}-\mathbf{a}_{0}e^{-s\sigma})^{-1}(\mathbf{b}+\mathbf{b}_{0}e^{-s\sigma}), 
    \\
    T_{0,N+1}(s)&= \mathbf{c}(s\mathbf{I}-\mathbf{a}-\mathbf{a}_{0}e^{-s\sigma})^{-1}\mathbf{b}_{0,N+1}e^{-s\sigma}, 
    \\
    T_{i,i+1}(s)&= \mathbf{c}(s\mathbf{I}-\mathbf{a}-\mathbf{a}_{\rm h}e^{-s\tau})^{-1}(\mathbf{b}+\mathbf{b}_{\rm h}e^{-s\tau}), 
    \\
    T_{N+1,0}(s)&= \mathbf{c}(s\mathbf{I}-\mathbf{a}-\mathbf{a}_{N+1}e^{-s\sigma})^{-1}\mathbf{b}_{N+1,0}e^{-s\sigma}, 
    \\
    T_{N+1,N+2}(s)&= \mathbf{c}(s\mathbf{I}-\mathbf{a}-\mathbf{a}_{N+1}e^{-s\sigma})^{-1}(\mathbf{b}+\mathbf{b}_{N+1}e^{-s\sigma}),
\end{split}
\end{align}
for ${i \in \{1, \ldots, N\}}$.
These link transfer functions can be calculated by substituting the coefficient matrices in~(\ref{eq:linmat}), and their expressions can be found in~(\ref{eq:seca-link-tf}) in the Appendix.

Using the link transfer functions, we can describe the overall response of the vehicle packet from vehicle ${N+2}$ to vehicle $0$ via the {\em head-to-tail transfer function}~\cite{Zhang2016}, $G_{0,N+2}$:
\begin{equation} \label{eq:AnHAG0-new}
\begin{split}
    \Tilde{V}_{0}(s)=G_{0,N+2}(s)\Tilde{V}_{N+2}(s),
\end{split}
\end{equation}
which is expressed from~(\ref{eq:responses}) as:
\begin{equation}\label{eq:head-to-tail-tf}
    G_{0,N+2}(s)
    =\frac{\big( T_{0,1}(s)T_{1,N+1}(s)\!+\!T_{0,N+1}(s) \big) T_{N+1,N+2}(s)}
    {1\!-\!\big( T_{0,1}(s)T_{1,N+1}(s)\!+\!T_{0,N+1}(s) \big) T_{N+1,0}(s)}.
\end{equation}
By following~\cite{Zhang2016}, this head-to-tail transfer function can be directly used for linear stability analysis.

%%%%%%%%%%%%%%%%%%%%%%%%%%%%%%%%%%%%%%%%%%%%%%%%%%%%%%%%%
\subsection{Stability}
%%%%%%%%%%%%%%%%%%%%%%%%%%%%%%%%%%%%%%%%%%%%%%%%%%%%%%%%%

In order to design the controllers of the CAVs, we formalize stability conditions for the vehicle packet through the notions of plant stability and string stability~\cite{Zhang2016, Feng2019}, using the head-to-tail transfer function.
Ultimately, this leads to the construction of {\em stability charts} that identify the controller parameters associated with plant and string stable vehicle packets, and hence guide the selection of these parameters.

{\em Plant stability} indicates that each vehicle in the fleet is able to approach the equilibrium state.
This is a fundamental requirement from CAVs to be operational in practice.
We analyze this by considering the characteristic equation:
\begin{equation}
    {\rm D}(G_{0,N+2}(s)) = 0,
\end{equation}
where ${\rm D}(.)$ denotes the denominator.
We denote the characteristic roots satisfying this equation by $s_{k}$ with ${k \in \mathbb{N}}$. The plant stability condition is established as $\operatorname{Re}(s_{k})<0$, ${\forall k \in \mathbb{N}}$, i.e., all characteristic roots must have negative real parts.
The system is at the plant stability boundary if either a real root ${s=0}$ is located at the imaginary axis, satisfying:
\begin{align} \label{eq:AnHAHPS0}
    {\rm D}(G_{0,N+2}(0))=0,
\end{align}
or a complex conjugate pair of roots ${s=\pm {\rm j}\Omega}$, with ${{\rm j}^2=-1}$ and some ${\Omega>0}$, is located at the imaginary axis, satisfying:
\begin{align}
\begin{split}
    \operatorname{Re}({\rm D}(G_{0,N+2}({\rm j} \Omega)))=0,
    \\
    \operatorname{Im}({\rm D}(G_{0,N+2}({\rm j} \Omega)))=0.
\end{split}
\label{eq:AnHAHPS1}
\end{align}

{\em String stability} indicates that speed perturbations are attenuated as they propagate upstream along the traffic.
This helps to avoid traffic congestion caused by growing speed perturbations on highways.
Specifically, we rely on the notion of \emph{head-to-tail string stability} \cite{Zhang2016}, wherein the speed fluctuation of the tail vehicle $|\Tilde{V}_{0}({\rm j}\omega)|$ is smaller than that of the lead vehicle $|\Tilde{V}_{N+2}({\rm j}\omega)|$ at any given frequency ${\omega>0}$.
Therefore, the string stability condition is established as:
\begin{equation} \label{eq:AnHAHSS}
    |G_{0,N+2}({\rm j}\omega)|<1,\quad\forall \omega > 0. %\geq
\end{equation}
In fact, this can be stated equivalently as ${P(\omega)>0}$ with:
\begin{equation} \label{eq:SS0P}
    P(\omega)\coloneqq \frac{1}{\omega^{2}} \! \left( {\rm D}(|G_{0,N+2}({\rm j}\omega)|^{2}) \!-\! {\rm N}(|G_{0,N+2}({\rm j}\omega)|^{2}) \right),
\end{equation}
where ${\rm D}(.)$ and ${\rm N}(.)$ denote denominator and numerator.

For the string stability boundaries, we consider two cases: ${\omega=0}$ and ${\omega>0}$.
For ${\omega=0}$, the boundaries are obtained by:
\begin{equation}
    P(0)=0,
    \label{eq:string_stab_om0}
\end{equation}
with applications of L'H{\^{o}}pital's rule to obtain $P(0)$ as the ${\omega \to 0}$ limit.
For ${\omega > 0}$, a family of string stability boundaries, parameterized by the wave number ${K\in [0,2\pi)}$, is given by:
\begin{equation} \label{eq:G_frac}
    G_{0,N+2}({\rm j}\omega)=e^{-{\rm j}K},
\end{equation}
see \cite{Molnar2022ch}.
Here one may write $G_{0,N+2}({\rm j}\omega)$ as the fraction:
\begin{equation} \label{eq:G_frac_expd}
    G_{0,N+2}({\rm j}\omega)\coloneqq \frac{a_{0}(\omega) + {\rm j} b_{0}(\omega)}{a_{1}(\omega) + {\rm j} b_{1}(\omega)},
\end{equation}
in which $a_{0}(\omega)$ and $b_{0}(\omega)$ are the real and imaginary parts of ${\rm N}(G_{0,N+2}({\rm j}\omega))$ while $a_{1}(\omega)$ and $b_{1}(\omega)$ are those of ${\rm D}(G_{0,N+2}({\rm j}\omega))$.
Then, (\ref{eq:G_frac}) can be decomposed into real and imaginary parts and rearranged to:
\begin{equation} \label{eq:SSwP}
\begin{split}
    a_{1}(\omega) - a_{0}(\omega) \cos K + b_{0}(\omega) \sin K&=0, \\
    b_{1}(\omega) - a_{0}(\omega) \sin K - b_{0}(\omega) \cos K&=0. \\
\end{split}
\end{equation}

To summarize, the plant stability boundaries are defined by~(\ref{eq:AnHAHPS0}) and~(\ref{eq:AnHAHPS1}), whereas the string stability boundaries are given by~(\ref{eq:string_stab_om0}) and~(\ref{eq:SSwP}).
These equations depend on the controller parameters, such as $\beta_{0,N+1}$ and $\beta_{N+1,0}$.
Thus, one may express these parameters and depict the stability boundaries in the $(\beta_{0,N+1},\beta_{N+1,0})$ plane.
The boundaries obtained from~(\ref{eq:AnHAHPS0}) and~(\ref{eq:string_stab_om0}) are of the form
${\beta_{N+1,0}=f_{0}(\beta_{0,N+1})}$,
the boundary from~(\ref{eq:AnHAHPS1}) is a curve parameterized by $\Omega$ in the form
${\beta_{0,N+1}=f_{1}(\Omega)}$, ${\beta_{N+1,0}=f_{2}(\Omega)}$,
whereas the boundaries from~(\ref{eq:SSwP}) are a family of curves parameterized by $\omega$ and $K$ as
${\beta_{0,N+1}=f_{3}(\omega,K)}$, ${\beta_{N+1,0}=f_{4}(\omega,K)}$.
The specific expressions of these boundaries can be found in the Appendix.

Depicting the stability boundaries leads to {\em stability charts} as the end result of the analysis.
The stability charts identify the regions of controller parameters that yield plant and string stable vehicle packet, so that these parameters can be selected as stabilizing control design.
We illustrate such stability charts in the next section for representative cases.

%%%%%%%%%%%%%%%%%%%%%%%%%%%%%%%%%%%%%%%%%%%%%%%
\begin{figure}[!t]
    \centering
    \includegraphics[scale=1]{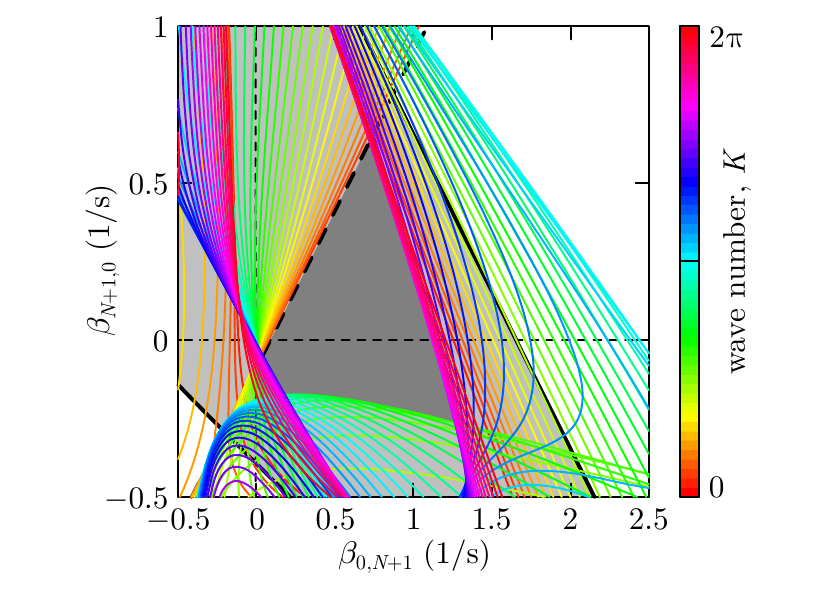}
     \caption{Stability chart of the scenario where a pair of CAVs encloses ${N=4}$ HVs.
     The light gray region enclosed by thick black curves is plant stable, and the dark gray region encapsulated by all curves is plant and string stable.}
	\label{fig:stability_chart}
\end{figure}
%%%%%%%%%%%%%%%%%%%%%%%%%%%%%%%%%%%%%%%%%%%%%%%

%%%%%%%%%%%%%%%%%%%%%%%%%%%%%%%%%%%%%%%%%%%%%%%%%%%%%%%%%
\section{Results}\label{sec:casestudy}
%%%%%%%%%%%%%%%%%%%%%%%%%%%%%%%%%%%%%%%%%%%%%%%%%%%%%%%%%

In this section, we present the results using stability charts of the linearized system~(\ref{eq:linearized_dynamics}) and simulations of the nonlinear system~(\ref{eq:CAV_model}) and~(\ref{eq:HV_model}).
We study the stability of mixed traffic that involves the CAV pair with the proposed controllers.
We demonstrate that information from connectivity is highly beneficial, and we analyze how the penetration of CAVs affects stability and the compactness of stable traffic.
The parameters of this case study are in Table~\ref{tab:parameters}.
Note that a worst-case scenario is considered in the sense that each HV is string unstable and significantly amplifies speed perturbations.

%%%%%%%%%%%%%%%%%%%%%%%%%%%%%%%%%%%%%%%%%%%%%%%%%%%%%%%%%
\subsection{Stability Charts}\label{sec:charts}
%%%%%%%%%%%%%%%%%%%%%%%%%%%%%%%%%%%%%%%%%%%%%%%%%%%%%%%%%

We show the stability boundaries obtained in the previous section by visualizing them as stability charts.
Fig.~\ref{fig:stability_chart} shows the stability chart in the ${(\beta_{0,N+1},\beta_{N+1,0})}$ plane for a vehicle packet with four HVs (${N=4}$).
While the ${s=0}$ plant stability boundary does not show up, the ${s=\pm {\rm j}\Omega}$ plant stability boundary is indicated by thick solid black line, and the plant stable region is shaded light gray.
Thick dashed black line shows the ${\omega = 0}$ string stability boundary, and thin curves in color denote the ${\omega>0}$ string stability boundaries for various values of ${K\in [0,2\pi)}$.
These curves bound the plant and head-to-tail string stable region in dark gray.
The control gains shall be selected from this region to achieve stability.

%%%%%%%%%%%%%%%%%%%%%%%%%%%%%%%%%%%%%%%%%%%%%%%
\begin{figure}
    \centering
    \includegraphics[scale=1]{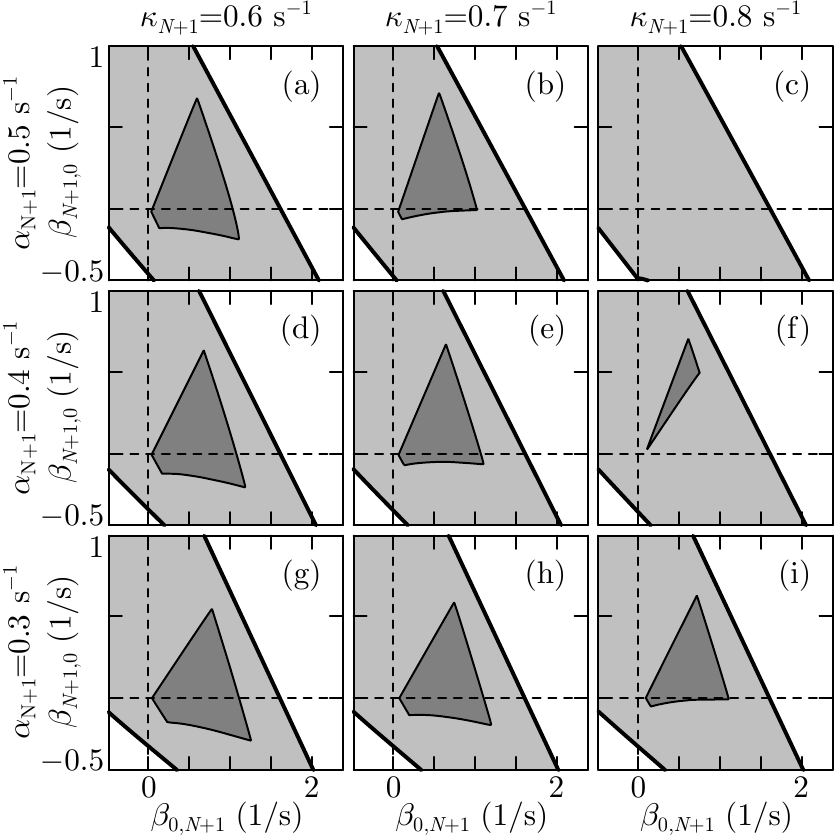}
     \caption{Stability charts of the scenario where a pair of CAVs encloses ${N=4}$ HVs, with various $\alpha_{N+1}$ and $\kappa_{N+1}$ parameter combinations.
     The same shading scheme is used as in Fig.~\ref{fig:stability_chart}.}
	\label{fig:stability_charts_alpha}
\end{figure}
%%%%%%%%%%%%%%%%%%%%%%%%%%%%%%%%%%%%%%%%%%%%%%%

Fig.~\ref{fig:stability_charts_alpha} shows stability charts for various headway response gains $\alpha_{N+1}$ and range policy gradients $\kappa_{N+1}$ of the head CAV.
The representation of the charts is simplified so that the boundary of the dark gray plant and string stable region is shown by a single thin black line; cf.~panel (d) that matches the case of Fig.~\ref{fig:stability_chart}.
Parameter $\alpha_{N+1}$ increases from bottom to top, while $\kappa_{N+1}$ increases from left to right across the panels.
As $\kappa_{N+1}$ increases the stable region shrinks significantly for larger $\alpha_{N+1}$, while it is less sensitive for smaller $\alpha_{N+1}$.
It is important to note that ${\beta_{0,N+1} \neq 0}$ and ${\beta_{0,N+1} \neq 0}$ are required for stability for certain $\alpha_{N+1}$ and $\kappa_{N+1}$; cf.~panel (f).
This indicates that response to the information from connectivity (associated with $\beta_{0,N+1}$ and $\beta_{N+1,0}$) is essential for stability.

\begin{figure}
    \centering
    \includegraphics[scale=1]{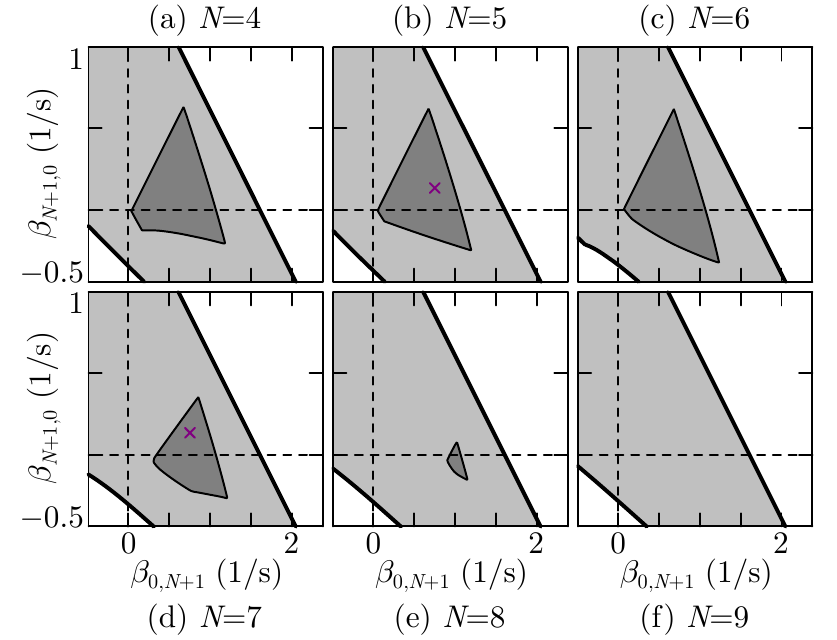}
     \caption{Stability chart of the scenario where a pair of CAVs encloses various numbers of HVs.
     The same shading scheme is used as in Fig.~\ref{fig:stability_chart}.}
	\label{fig:stability_charts_N}
\end{figure}

Importantly, the number $N$ of HVs between the CAV pair also affects the stability charts.
Fig.~\ref{fig:stability_charts_N} shows stability charts for
vehicle packets with various numbers of HVs: ${N=4,5,6,7,8,9}$ (where panel (a) matches Fig.~\ref{fig:stability_chart}).
Since the human
driver parameters are selected to be string unstable, the head-to-tail string stable region shrinks
as the number $N$ of HVs increases.
Yet, we can establish plant and string stable traffic control for up to ${N=8}$ HVs by the right choice of $\beta_{0,N+1}$ and $\beta_{N+1,0}$.
Moreover, even when the setup is head-to-tail string unstable (${N=9}$), the proposed controller may help to mitigate the instability.
Notice that in practice it may be difficult for the CAVs to identify the number $N$ of non-connected HVs between them.
Thus, when designing the CAVs' controllers in~(\ref{eq:controller}) it is preferable to choose gains that provide string stability robustly for a range of $N$ (i.e., gains that lie in the intersection of the stable domains calculated for various $N$). 
For example, the gain combinations corresponding to the  purple cross ensure string stability for ${N=4,5,6,7}$.
Alternatively, one may consider the aspects of energy efficiency~\cite{He2020} or robustness w.r.t. human driver behavior~\cite{Hajdu2020} when selecting gains from the stable domain.

%%%%%%%%%%%%%%%%%%%%%%%%%%%%%%%%%%%%%%%%%%%%%%%
%%%%%%%%%%%%%%%%%%%%%%%%%%%%%%%%%%%%%%%%%%%%%%%

%%%%%%%%%%%%%%%%%%%%%%%%%%%%%%%%%%%%%%%%%%%%%%%%%%%%%%%%%
\subsection{CAV Penetration and the Compactness of Stable Traffic }
%%%%%%%%%%%%%%%%%%%%%%%%%%%%%%%%%%%%%%%%%%%%%%%%%%%%%%%%%

According to the example in Fig.~\ref{fig:stability_charts_N}, the minimum penetration of connectivity and automation required for a stable vehicle packet is $20\%$ (the CAV pair can stabilize the packet with at most ${N=8}$ HVs).
While this penetration may seem high, it is important to recall that our case study involves a worst-case scenario where every HV behaves string unstable and may significantly amplify speed perturbations (i.e., ${|T_{i,i+1}({\rm j}\omega)|>1}$ at some ${\omega>0}$, with maximum ${|T_{i,i+1}(0.58{\rm j})| \approx 1.03}$).
We also remark that the typical range of connectivity is a few hundred meters~\cite{Molnar2022acc}.
Thus, it may not always be feasible to connect across more than ${N=8}$ HVs.
As such, the proposed strategy is able to stabilize even the worst-case vehicle packets that the CAV pair may connect across.

Motivated by this example, now we dive deeper into how the CAV penetration affects the stability and compactness of traffic.
We quantify a fundamental trade-off: more compact traffic requires higher penetration of CAVs to maintain stability.
First, we define the {\em penetration} by:
\begin{equation}\label{eq:penetration}
    p = \frac{2}{N+2},
\end{equation}
since two CAVs form a vehicle packet with $N$ number of HVs.
Furthermore, to characterize {\em traffic compactness}, we define the average equilibrium headway of the vehicle packet:
\begin{equation}\label{eq:average_headway}
    \bar{h}=\frac{h^{*}_{0}+N h_{\rm h}^{*}+h^{*}_{N+1}}{N+2}.
\end{equation}

We seek to identify the relation between the penetration $p$ and the average headway $\bar{h}$ such that the vehicle packet can be maintained plant and string stable by the CAV pair.
To achieve this goal, we vary the number $N$ of HVs and parameter $\kappa_{N+1}$ associated with both the head-to-tail string stability as well as with the equilibrium headway ${h_{N+1}^{*}=v^{*}/\kappa_{N+1} + h_{\mathrm{st}}}$ ahead of the vehicle packet; cf.~\eqref{eq:kappas_vs_h}.
We plot stability charts for each $(N,\kappa_{N+1})$ parameter combination (similar to Figs.~\ref{fig:stability_charts_alpha} and~\ref{fig:stability_charts_N}), and we identify the maximum gradient $\kappa_{N+1,\max}$ for which stable region exists as a function of $N$.
Finally, we convert this value to the minimum average headway $\bar{h}_{\min}$ as a function of the penetration $p$ using~\eqref{eq:kappas_vs_h},~\eqref{eq:penetration} and~\eqref{eq:average_headway}:
\begin{equation}
    \bar{h}_{\min}= \frac{p}{2} \bigg( h^{*}_{0} + \frac{v^{*}}{ \kappa_{N+1,\max}} + h_{\mathrm{st}} \bigg)+ (1-p) h_{\rm h}^{*},
\end{equation}
where $\kappa_{N+1,\max}$ depends on $p$. We evaluate this formula for ${v^{*}=20\,{\rm m/s}}$.

%%%%%%%%%%%%%%%%%%%%%%%%%%%%%%%%%%%%%%%%%%%%%%%
\begin{figure}
    \centering
    \includegraphics[scale=1]{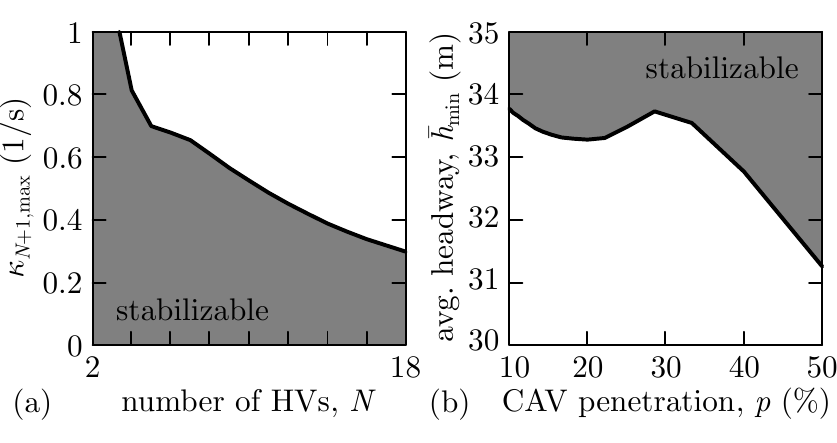}
     \caption{Relationship between the compactness of stabilizable traffic and the penetration of connectivity and automation.
     (a) The maximum range policy gradient $\kappa_{N+1,\max}$ of the head CAV that allows stable traffic as a function of the number $N$ of HVs between the two CAVs.
     (b) The minimum average headway $\bar{h}_{\min} $ that allows stable traffic as a function of the penetration $p$ of CAVs.
     There is a fundamental trade-off: for smaller penetration of CAVs one needs higher average headway (less compact traffic) to be able to maintain stability.}
	\label{fig:penetration_stability}
\end{figure}
%%%%%%%%%%%%%%%%%%%%%%%%%%%%%%%%%%%%%%%%%%%%%%%

The resulting penetration versus average headway diagram is shown in Fig.~\ref{fig:penetration_stability}.
It quantifies the trade-off between CAV penetration, traffic compactness and stability: one typically requires higher penetration of connectivity and automation to achieve more compact stable traffic.
Equivalently, one usually needs higher average headway to stabilize traffic with lower penetration of CAVs.
For example, 10\% CAV penetration requires about 34 m average headway for stability while 50\% requires 31 m.
Note that the trend is not fully monotonous, and there is a local maximum on Fig.~\ref{fig:penetration_stability}(b) around ${p=30\%}$.

%%%%%%%%%%%%%%%%%%%%%%%%%%%%%%%%%%%%%%%%%%%%%%%%%%%%%%%%%
\subsection{Simulations}
%%%%%%%%%%%%%%%%%%%%%%%%%%%%%%%%%%%%%%%%%%%%%%%%%%%%%%%%%

Finally, we present numerical simulation results for the nonlinear system~(\ref{eq:CAV_model}) and~(\ref{eq:HV_model}) to investigate the performance and robustness of the CAVs' controllers.
The underlying parameters are listed in Table~\ref{tab:parameters} and at the figures.

%%%%%%%%%%%%%%%%%%%%%%%%%%%%%%%%%%%%%%%%%%%%%%%
\begin{figure}[t]
    \centering
    \includegraphics[scale=1]{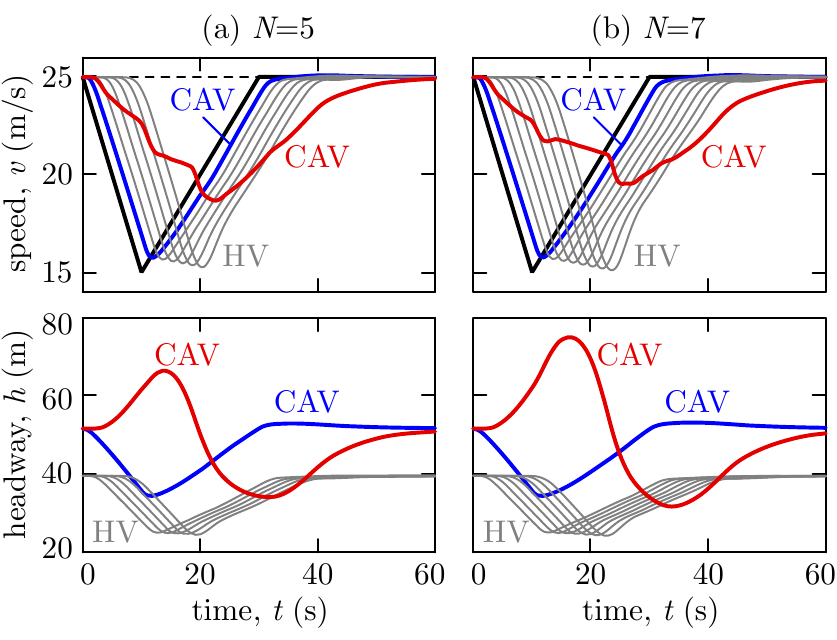}
     \caption{Simulation of the vehicle packet that shows the robustness of the CAV pair's controller against the number $N$ of HVs, with (a) ${N=5}$ and (b) ${N=7}$.
     The controller parameters correspond to the purple $\times$ in Fig.~\ref{fig:stability_charts_N}(b,d).}
	\label{fig:simulation}
\end{figure}
%%%%%%%%%%%%%%%%%%%%%%%%%%%%%%%%%%%%%%%%%%%%%%%

%%%%%%%%%%%%%%%%%%%%%%%%%%%%%%%%%%%%%%%%%%%%%%%
\begin{figure*}[t]
	\centering
    \includegraphics[scale=1.04]{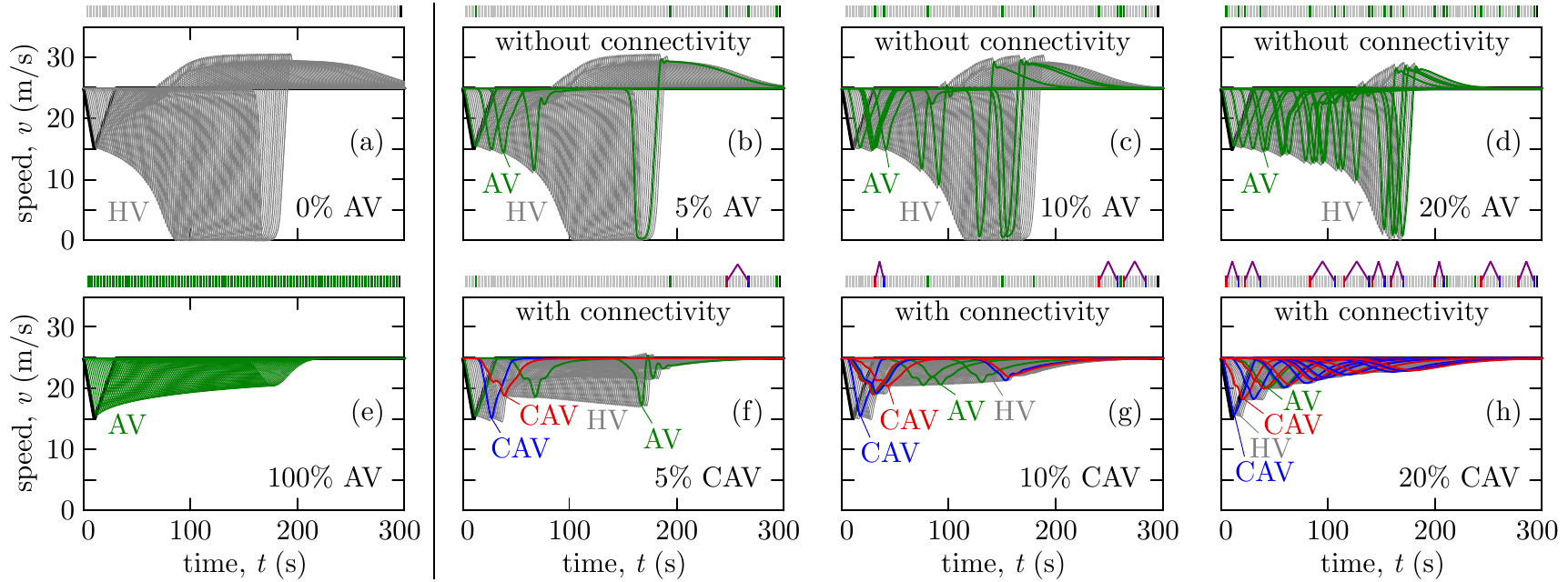}
    \caption{Simulation of mixed traffic including 100 vehicles with various penetrations of CAVs.
    (a) String unstable human-driven traffic.
    (e) 100\% penetration of non-connected AVs executing ACC.
    (b)-(d) Without connectivity, 5\%, 10\%, and 20\% penetrations of ACC-capable AVs are not able to mitigate the onset of a traffic congestion.
    (f)-(h) With connectivity, 5\%, 10\%, and 20\% of CAVs are able to mitigate the congestion by forming pairs and executing the proposed control strategy.
    Note that connectivity (panels (g) and (h)) even yields smaller speed fluctuations than 100\% non-connected ACC-capable AVs (panel (e)).}
    \label{fig:simulation_large_scale}
\vspace{-2mm}
\end{figure*}
%%%%%%%%%%%%%%%%%%%%%%%%%%%%%%%%%%%%%%%%%%%%%%%

First, we simulate a single vehicle packet.
Fig.~\ref{fig:simulation} shows simulation results considering the lead vehicle motion from Fig.~\ref{fig:benefits}.
Since the number $N$ of HVs in the packet may be unknown to the CAVs in practice, the robustness of the controller with respect to $N$ is demonstrated in the figure.
The same controller (with the same gains) is tested on setups with different $N$: ${N=5}$ (left) and ${N=7}$ (right).
In both cases, the CAV pair successfully achieves head-to-tail string stability -- the tail CAV reduces its speed less than the lead vehicle -- while maintaining safe headways, with similar motion.
In this particular example, the tail CAV plays a bigger role in attenuating velocity fluctuations than the head CAV.
Alternatively, string stability could also be achieved through the smooth driving of the head CAV, which was shown to be a successful approach for a single leading CAV in~\cite{Molnar2020cdc, Molnar2022ch, wang2020lcc, wang2022deeplcc}. Utilizing such approach can further develop the potential of attenuating velocity fluctuations by the CAV pair.

Second, we study large-scale mixed traffic that consists of HVs and CAVs, 100 vehicles in total.
We fix the number (penetration) of CAVs and distribute them randomly in traffic.
We form CAV pairs by going through the 100 vehicles from head to tail.
Whenever two CAVs have ${1 \leq N \leq 7}$ HVs between them, they form a pair, and we move onto the next unpaired CAV.
If a CAV does not have other CAVs nearby (${N>7}$) or follows another CAV (${N=0}$), it is labelled as AV and commanded to execute adaptive cruise control (as the ${\beta_{0,N+1}=\beta_{N+1,0}=0}$ special case of~(\ref{eq:controller})).
Furthermore, we make comparison with a no-connectivity baseline where all CAVs are left unpaired and execute ACC as AVs.

Fig.~\ref{fig:simulation_large_scale} shows simulation results for $0\%$, $5\%$, $10\%$, $20\%$, and $100\%$ penetration of CAVs.
The distribution of the different vehicle types in traffic is depicted at the top of each panel.
Gray color indicates HVs, green shows (non-connected) AVs, while blue and red denote CAV pairs.
Fig.~\ref{fig:simulation_large_scale}(a) shows the $0\%$ penetration reference case of human-driven traffic, which exhibits a stop-and-go congestion since HVs are string unstable.
This highlights the challenge for CAVs to stabilize traffic.
As opposed, the $100\%$ penetration baseline in Fig.~\ref{fig:simulation_large_scale}(e), where traffic consists of ACC-capable AVs only, mitigates the congestion through string stable behavior.
While this scenario is ideal, it is achieved by extremely large AV penetration.

Fig.~\ref{fig:simulation_large_scale}(b)-(d) and~(f)-(h) present more realistic, $5\%$, $10\%$ and $20\%$ penetrations.
Cases without connectivity (top) and with connectivity (bottom) are compared, where CAVs act as ACC-capable AVs and where nearby CAVs are paired, respectively.
The figure clearly shows that the proposed CAV pairs (bottom) significantly improve traffic smoothness compared to the no-connectivity baseline (top).
Ultimately, string stability is achieved at low, $10\%$ CAV penetration, and speed fluctuations further decrease as penetration increases (bottom).
This cannot be achieved with low penetration of non-connected AVs (top).
Moreover, low penetration of CAVs even outperforms the full penetration of AVs without connectivity; cf.~Figs.~\ref{fig:simulation_large_scale}(e) and~(h).

We remark that controller~(\ref{eq:controller}) does not have formal guarantees of maintaining safe distances ahead of the CAVs.
To remedy this, we tuned the controller such that the headway of each simulated vehicle was positive and no collision occurred.
We will seek to address guaranteed safety in our future work.
Furthermore, to avoid vehicles moving in reverse (${v_i(t)<0}$), we modified the input of each simulated vehicle (both HVs and CAVs) to ${\hat{u}_i(t) = \max\{u_i(t),-\alpha_{v}v_i(t)}\}$ with ${\alpha_v = 10\,{\rm s^{-1}}}$.
This affected the results close to zero speed only.
Finally, note that the lead vehicle motion from Fig.~\ref{fig:benefits} was considered in the simulation results.
By exploring various lead vehicle motions, we noticed the occurrence of a bistability phenomenon in which speed perturbations decay for certain lead vehicle motions but amplify for others.
Studying the effect of bistability is left for future work, and we restrict ourselves to the specific lead vehicle motion from Fig.~\ref{fig:benefits}.

We further study the string stability of the 100-vehicle traffic by quantifying the maximum speed fluctuation of each vehicle relative to the lead vehicle, by introducing:
\begin{equation}
    \Gamma_i = \frac{\max_{t \geq 0} |v_{i}(t)-v_{i}(0)|}{\max_{t \geq 0} |v_{100}(t)-v_{100}(0)|}, \quad
    \bar{\Gamma} = \frac{1}{100} \sum_{i=0}^{99} \Gamma_{i},
\end{equation}
where ${\Gamma_{0} < 1}$ implies head-to-tail string stability, while $\bar{\Gamma}$ is an average stability metric across all 100 vehicles.

%%%%%%%%%%%%%%%%%%%%%%%%%%%%%%%%%%%%%%%%%%%%%%%
\begin{figure}
    \centering
    \includegraphics[scale=1.1]{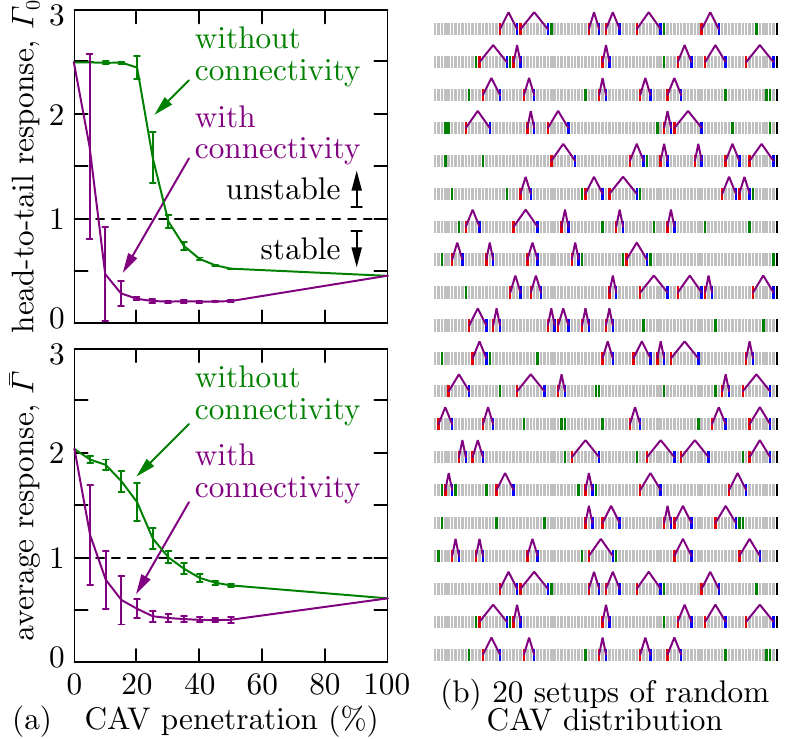}
     \caption{(a) Head-to-tail and average string stability metrics as a function of the CAV penetration.
     By exploiting connectivity, stability can be achieved at penetrations as low as 10\%, while without connectivity one requires 30\% penetration.
     (b) Illustration of the 20 simulation setups with 15\% penetration that were used to obtain the stability metrics.}
	\label{fig:penetration_response}
\vspace{-3mm}
\end{figure}
%%%%%%%%%%%%%%%%%%%%%%%%%%%%%%%%%%%%%%%%%%%%%%%

Fig.~\ref{fig:penetration_response}(a) shows the stability metrics $\Gamma_{0}$ and $\bar{\Gamma}$ as a function of the CAV penetration.
For each penetration, 20 setups are simulated in which the positions (indices) of CAVs in traffic are allocated randomly, as illustrated for 15\% penetration in Fig.~\ref{fig:penetration_response}(b).
Fig.~\ref{fig:penetration_response}(a) shows the mean and standard deviation of the 20 simulation results.
When all vehicles are HVs ($0\%$ penetration), the 20 cases coincide, and most vehicles undergo speed fluctuations more than twice of the lead (${\bar{\Gamma}>2}$). 
With automation (nonzero penetration), two cases are compared: without connectivity (i.e., AVs executing ACC, cf.~the top of Fig.~\ref{fig:simulation_large_scale}) and with connectivity (i.e., CAV pairs, cf.~the bottom of Fig.~\ref{fig:simulation_large_scale}).
With connectivity, head-to-tail string stability ($\Gamma_{0}<1$) is achieved at a minimum penetration of $10\%$. 
As penetration increases, speed fluctuations further decrease to ${\bar{\Gamma} \approx 0.4}$.
In comparison, automation without connectivity achieves head-to-tail string stability only at penetrations around $30\%$ and speed perturbations only decrease to ${\bar{\Gamma} \approx 0.6}$.

Finally, we remark that if the penetration of CAVs is large enough, then a given CAV may travel within the communication range of and could respond to multiple other CAVs.
This could lead to more complex CAV networks than the CAV pair setup.
Note that the problem of creating groups between multiple CAVs is well studied in the literature, especially in the context of truck platooning~\cite{VanDeHoef2020, Jin2021}.
As such, the control law~(\ref{eq:controller}) is scalable to include the interaction of more that two CAVs.
For example, one may consider the extension:
\begin{equation}
    u_{i} =\alpha_{i}\big(V_{i}(h_{i})-v_{i}\big) + \beta_{i}\big(W(v_{i})-v_{i}\big) + \sum_{j \in \mathcal{J}_i} \beta_{i,j}(W(v_{j})-v_{i})
\end{equation}
where ${\mathcal{J}_i}$ is the set of CAVs that are in the communication range of CAV $i$.
The analysis of such larger CAV networks, however, is more involved, since the number of tunable control gains increases and the expression~(\ref{eq:head-to-tail-tf}) of the head-to-tail transfer function becomes more complex.
To keep our exposition simple, we presented the CAV pair setup only, and excluded further CAV connections from the simulation examples too.
Considering general CAV networks could lead to further benefits, and hence it is left for future work.

%%%%%%%%%%%%%%%%%%%%%%%%%%%%%%%%%%%%%%%%%%%%%%%%%%%%%%%%%
\section{Conclusions}\label{sec:conclusions}
%%%%%%%%%%%%%%%%%%%%%%%%%%%%%%%%%%%%%%%%%%%%%%%%%%%%%%%%%

This work proposes {\em connected cruise and traffic control}, wherein {\em pairs of connected automated vehicles} (CAVs) regulate their longitudinal motion amongst human-driven vehicles to stabilize traffic, by exploiting vehicle-to-everything (V2X) connectivity.
Stability analysis and numerical simulations are conducted to characterize the performance of the CAV pair.
It is shown that the CAV pair is able to significantly improve the smoothness of traffic flow, which resembles the phenomenon of {\em Cooper pairs} \cite{Cooper1956} in superconductors.
The trade-off between the required CAV penetration and the compactness of stabilizable traffic is quantified via stability charts.
Moreover, large-scale traffic simulations show the impact of CAV penetration on traffic smoothness and the benefits of connectivity.
Potential future work may include exploring the energy efficiency of the CAV pair setup and ensuring formal safety guarantees via control barrier functions.

%%%%%%%%%%%%%%%%%%%%%%%%%%%%%%%%%%%%%%%%%%%%%%%%%%%%%%%%%
\appendices
\section*{Appendix}
%%%%%%%%%%%%%%%%%%%%%%%%%%%%%%%%%%%%%%%%%%%%%%%%%%%%%%%%%

Here we present the technical details related to the stability analysis of Section~\ref{sec:stability}, including the expressions of the link transfer functions, and plant and string stability boundaries for ${N>0}$.
The resulting formulas~(\ref{eq:boundary_plant}),~(\ref{eq:boundary_string_1}) and~(\ref{eq:boundary_string_2}) were used to plot the stability charts in Section~\ref{sec:charts}.

First, we introduce the following combined parameters:
\begin{equation}
    \xi_{i} = \alpha_{i} \kappa_{i}, \quad
    \eta_{i} = \alpha_{i} + \beta_{i}, \quad
    \zeta_{i} = \alpha_{i}+2\beta_{i}-2\kappa_{i},
\end{equation}
${i \in \{0, N+1, {\rm h}\}}$.
Then, the coefficient matrices in~(\ref{eq:linearized_dynamics}) are:
\begin{equation} \label{eq:linmat}
\begin{split}
    \mathbf{a}=& \begin{bmatrix} 
    0 & -1 \\ 0 & 0  
    \end{bmatrix}\!,\;\;
    \qquad\,\mathbf{a}_{0}=\begin{bmatrix}
        0 & 0 \\ \xi_{0} & -(\eta_{0}+\beta_{0,N+1}) 
    \end{bmatrix}\!, 
    \\
    \mathbf{a}_{\rm h}=& \begin{bmatrix}
        0 & 0 \\ \xi_{\rm h}  & -\eta_{\rm h}
    \end{bmatrix}\!,\;\;
    \mathbf{a}_{N+1}= \begin{bmatrix}
        0 & 0 \\ \xi_{N+1}  & -(\eta_{N+1}+\beta_{N+1,0}) 
    \end{bmatrix}\!,
    \\
    \mathbf{b}=& \begin{bmatrix}
        1 \\ 0
    \end{bmatrix}\!,\;\;
    \qquad\mathbf{b}_{0}= \begin{bmatrix}
        0 \\ \beta_{0}
    \end{bmatrix}\!,\;\;
    \quad\,\mathbf{b}_{0,N+1}= \begin{bmatrix}
        0 \\ \beta_{0,N+1}
    \end{bmatrix},\\
    \mathbf{b}_{\rm h}=& \begin{bmatrix}
        0 \\ \beta_{\rm h}
    \end{bmatrix}\!,\;\;
    \mathbf{b}_{N+1}= \begin{bmatrix}
        0 \\ \beta_{N+1}
    \end{bmatrix}\!,\;\;
    \mathbf{b}_{N+1,0}= \begin{bmatrix}
        0 \\ \beta_{N+1,0}
    \end{bmatrix}\!.
    \end{split}
\end{equation}
By substituting (\ref{eq:linmat}) into (\ref{eq:sec3a-link-tf}), the link transfer functions read:
\begin{equation} \label{eq:seca-link-tf}
    \begin{split}
        T_{0,1}(s)&= \frac{\beta_{0}s+\xi_{0}}{s^2 e^{s\sigma}+(\eta_{0}+\beta_{0,N+1})s+\xi_{0}}, 
        \\
        T_{0,N+1}(s)&= \frac{\beta_{0,N+1}s}{s^2 e^{s\sigma}+(\eta_{0}+\beta_{0,N+1})s+\xi_{0}}, 
        \\
        T_{i,i+1}(s) &= \frac{\beta_{\rm h}s+\xi_{\rm h}}{s^{2}e^{s\tau}+\eta_{\rm h}s+\xi_{\rm h}}, 
        \\
        T_{N+1,N+2}(s)&= \frac{\beta_{N+1} s+\xi_{N+1}}{s^{2} e^{s\sigma}+(\eta_{N+1}+\beta_{N+1,0})s+\xi_{N+1}}, 
        \\
        T_{N+1,0}(s)&= \frac{\beta_{N+1,0}s}{s^{2} e^{s\sigma}+(\eta_{N+1}+\beta_{N+1,0})s+\xi_{N+1}}.
    \end{split}
\end{equation}

The ${s=0}$ plant stability boundary is given by~(\ref{eq:AnHAHPS0}), that, after substituting~(\ref{eq:head-to-tail-tf}) and~(\ref{eq:seca-link-tf}), leads to ${\alpha_{N+1}=0}$ and ${\alpha_{0}=0}$.
These boundaries do not appear in the $(\beta_{0,N+1},\beta_{N+1,0})$ plane.
The ${s=\pm {\rm j}\Omega}$ plant stability boundary is given by~(\ref{eq:AnHAHPS1}).
By substituting $G_{0,N+2}({\rm j}\Omega)$ from~(\ref{eq:head-to-tail-tf}) and~(\ref{eq:seca-link-tf}) into~(\ref{eq:AnHAHPS1}), we get:
\begin{align}
\begin{split}
p_{1}(\Omega)\beta_{0,N+1}+q_{1}(\Omega)\beta_{N+1,0}+r_{1}(\Omega)=0, 
\\
p_{2}(\Omega)\beta_{0,N+1}+q_{2}(\Omega)\beta_{N+1,0}+r_{2}(\Omega)=0,
\end{split}
\end{align}
where the coefficients are:
\begin{align}
\begin{split}
    p_{1}(\Omega)&=\Omega^{3}S-\Omega^{2}\eta_{N+1},
    \\
    q_{1}(\Omega)&=\Omega^{3}S+\Omega \big( w_{1}(\Omega)-\Omega\eta_{0} \big),
    \\
    r_{1}(\Omega)&=\Omega^{4} (C^2-S^2)+\Omega^{3}(\eta_{N+1}+\eta_{0})S
    \\
    &\quad-\Omega^{2}(\xi_{N+1}+\xi_{0})C-\Omega^{2}\eta_{N+1}\eta_{0}+\xi_{N+1}\xi_{0},
    \\
    w_{1}(\Omega)&=\Omega\beta_{0}\Gamma_{\mathrm{R}}(\Omega) + \xi_{0}\Gamma_{\mathrm{I}}(\Omega),
    \\
    p_{2}(\Omega)&=-\Omega^{3}C+\Omega\xi_{N+1},
    \\
    q_{2}(\Omega)&=-\Omega^{3}C+\Omega \big( w_{2}(\Omega) + \xi_{0} \big),
    \\
    r_{2}(\Omega)&=2\Omega^{4}SC
    -\Omega^{3}(\eta_{N+1}+\eta_{0})C
    \\
    &\quad-\Omega^{2}(\xi_{N+1}+\xi_{0})S
    +\Omega \big( \xi_{0} \eta_{N+1} + \xi_{N+1} \eta_{0} \big),
    \\
    w_{2}(\Omega)&=\Omega\beta_{0}\Gamma_{\mathrm{I}}(\Omega)-\xi_{0}\Gamma_{\mathrm{R}}(\Omega),
\end{split}
\end{align}
with
${\Gamma_{\mathrm{R}}(\Omega) = \operatorname{Re}\big(T_{i,i+1}({\rm j}\Omega)^{N}\big)}$,
${\Gamma_{\mathrm{I}}(\Omega) = \operatorname{Im}\big(T_{i,i+1}({\rm j}\Omega)^{N}\big)}$,
${S = \sin{(\Omega\sigma)}}$,
${C = \cos{(\Omega\sigma)}}$.
Thus, the plant stability boundary becomes:
\begin{align}
\begin{split}
    \beta_{0,N+1}=\frac{q_{2}(\Omega)r_{1}(\Omega)-q_{1}(\Omega)r_{2}(\Omega)}{p_{2}(\Omega)q_{1}(\Omega)-p_{1}(\Omega)q_{2}(\Omega)}, 
    \\
    \beta_{N+1,0}=\frac{p_{1}(\Omega)r_{2}(\Omega)-p_{2}(\Omega)r_{1}(\Omega)}{p_{2}(\Omega)q_{1}(\Omega)-p_{1}(\Omega)q_{2}(\Omega)}.
\end{split}
\label{eq:boundary_plant}
\end{align}

The ${\omega = 0}$ string stability boundary is defined by~(\ref{eq:SS0P}) and~(\ref{eq:string_stab_om0}).
By substituting~(\ref{eq:head-to-tail-tf}) and~(\ref{eq:seca-link-tf}) into~(\ref{eq:SS0P}), we obtain:
\begin{equation}\label{eq:P_expression}
\begin{split}
    &P(\omega)\!=\!\xi_{N+1}^{2} \xi_{0}^{2} \frac{1\!-\!\Gamma_{\mathrm{I}}^{2}(\omega)\!-\!\Gamma_{\mathrm{R}}^{2}(\omega)}{\omega^{2}}
    \!+\! \xi_{N+1}^{2} \alpha_{0} \zeta_{0}
    \!+\! \xi_{0}^{2} \alpha_{N+1} \zeta_{N+1} 
    \\
    &+\!2 \xi_{N+1} \alpha_{0} (\xi_{0} \beta_{N+1,0} \!-\! \xi_{N+1} \beta_{0,N+1}) \Big( \kappa_{0} \frac{\Gamma_{\mathrm{I}}(\omega)}{\omega} \!-\! 1 \Big)
    \!+\! \mathcal{O}(\omega).
\end{split}
\end{equation}
Note that ${\lim_{\omega\rightarrow 0}\Gamma_{\mathrm{R}}(\omega)=1}$ and ${\lim_{\omega\rightarrow 0}\Gamma_{\mathrm{I}}(\omega)=0}$ hold,
hence terms of $1-\Gamma_{\mathrm{R}}(\omega)$ and $\Gamma_{\mathrm{I}}(\omega)$
were absorbed into $\mathcal{O}(\omega)$.
Then, one may take the limit ${\omega\rightarrow 0}$ to find the ${\omega=0}$ string stability boundary via~(\ref{eq:string_stab_om0}).
By using the expression~(\ref{eq:seca-link-tf}) of $T_{i,i+1}$ and applying L'H{\^{o}}pital's rule, the following holds~\cite{Molnar2022ch}: 
\begin{equation}
    \lim_{\omega\rightarrow 0}\!\frac{\Gamma_{\mathrm{I}}(\omega)}{\omega}\!=\!-\frac{N}{\kappa_{\rm h}}, \;\
    \lim_{\omega\rightarrow 0}\!\frac{1\!-\!\Gamma_{\mathrm{I}}^{2}(\omega)\!-\!\Gamma_{\mathrm{R}}^{2}(\omega)}{\omega^{2}}\!=\!\frac{N\alpha_{\rm h}\zeta_{\rm h}}{\xi_{\rm h}^{2}}.
\end{equation}
This puts~(\ref{eq:string_stab_om0}) into the form:
\begin{equation}
    p \beta_{0,N+1} + q \beta_{N+1,0} + r = 0,
\end{equation}
in which the coefficients read:
\begin{align}
\begin{split}
    p&=2\xi_{N+1}^{2} \alpha_{0}\Big( 1+N \frac{\kappa_{0}}{\kappa_{\rm h}} \Big), \\
    q&=-p \frac{\xi_{0}}{\xi_{N+1}}, 
    \\
    r&=\frac{\xi_{N+1}^{2} \xi_{0}^{2}}{\xi_{\rm h}^{2}} N \alpha_{\rm h}\zeta_{\rm h}
    + \xi_{N+1}^{2} \alpha_{0} \zeta_{0}
    + \xi_{0}^{2} \alpha_{N+1} \zeta_{N+1}.
\end{split}
\end{align}
This finally leads to the ${\omega = 0}$ string stability boundary:
\begin{equation}\label{eq:boundary_string_1}
    \beta_{0,N+1} = -\frac{q}{p} \beta_{N+1,0} - \frac{r}{p}.
\end{equation}

The ${\omega > 0}$ string stability boundary is obtained from~(\ref{eq:SSwP}).
By using the expression of $G_{0,N+2}({\rm j}\omega)$ from~(\ref{eq:head-to-tail-tf}) and~(\ref{eq:seca-link-tf}), one gets $a_{0}(\omega)$ and $b_{0}(\omega)$ as the real and imaginary parts of ${\rm N}(G_{0,N+2}({\rm j}\omega))$, and one obtains $a_{1}(\omega)$ and $b_{1}(\omega)$ as those of ${\rm D}(G_{0,N+2}({\rm j}\omega))$.
After re-organizing~(\ref{eq:SSwP}), we have:
\begin{align}
\begin{split}
p_{1}'(\omega,K)\beta_{0,N+1}+q_{1}(\omega)\beta_{N+1,0}+r_{1}'(\omega,K)=0, 
\\
p_{2}'(\omega,K)\beta_{0,N+1}+q_{2}(\omega)\beta_{N+1,0}+r_{2}'(\omega,K)=0,
\end{split}
\end{align}
where:
\begin{align}
\begin{split}
    p_{1}'(\omega,K) &= \omega^{2}\beta_{N+1} \cos K + \omega\xi_{N+1} \sin K + p_1(\omega),
    \\
    r_{1}'(\omega,K) &= \big( \omega\beta_{N+1} w_{1}(\omega) + \xi_{N+1} w_{2}(\omega) \big) \cos K 
    \\
    & - \big( \omega\beta_{N+1} w_{2}(\omega) - \xi_{N+1} w_{1}(\omega) \big) \sin K + r_1(\omega),
    \\
    p_{2}'(\omega,K) &= \omega^{2}\beta_{N+1} \sin K - \omega\xi_{N+1} \cos K + p_2(\omega),
    \\
    r_{2}'(\omega,K) &= \big( \omega\beta_{N+1} w_{1}(\omega) + \xi_{N+1} w_{2}(\omega) \big) \sin K 
    \\
    & + \big( \omega\beta_{N+1} w_{2}(\omega) - \xi_{N+1} w_{1}(\omega) \big) \cos K + r_2(\omega).
\end{split}
\end{align}
Therefore, the string stability boundaries are solved as:
\begin{align}
    \begin{split}
    \beta_{0,N+1}&=\frac{q_{2}(\omega)r_{1}'(\omega,K)-q_{1}(\omega)r_{2}'(\omega,K)}{p_{2}'(\omega,K)q_{1}(\omega)-p_{1}'(\omega,K)q_{2}(\omega)}, 
    \\ 
    \beta_{N+1,0}&=\frac{p_{1}'(\omega,K)r_{2}'(\omega,K)-p_{2}'(\omega,K)r_{1}'(\omega,K)}{p_{2}'(\omega,K)q_{1}(\omega)-p_{1}'(\omega,K)q_{2}(\omega)}.
    \end{split}
    \label{eq:boundary_string_2}
\end{align}

\bibliographystyle{IEEEtran}
\bibliography{ref}

% \vspace{-9mm}
\begin{IEEEbiography}
[{\includegraphics[width=1in,height=1.25in,clip,keepaspectratio]{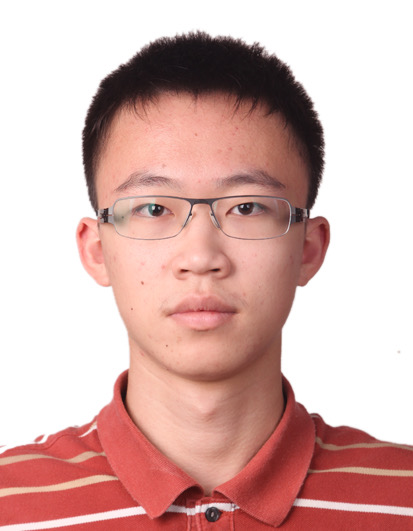}}]
{Sicong Guo} received his BSc and MSc degrees in Mechanical Engineering from the University of Michigan, Ann Arbor in 2020 and 2022, respectively. His research interests include the dynamics and control of time delay systems, with application to connected automated vehicles and robotic systems.
\end{IEEEbiography}

% \vspace{-9mm}
\begin{IEEEbiography}[{\includegraphics[width=1in,height=1.25in,clip,keepaspectratio]{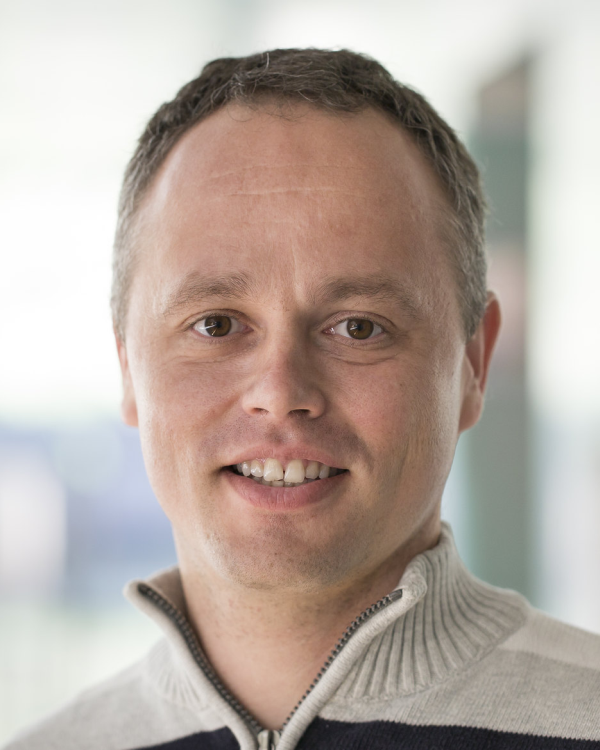}}]{G\'abor Orosz} received the MSc degree in Engineering Physics from the Budapest University of Technology, Hungary, in 2002 and the PhD degree in Engineering Mathematics from the University of Bristol, UK, in 2006. He held postdoctoral positions at the University of Exeter, UK, and at the University of California, Santa Barbara. In 2010, he joined the University of Michigan, Ann Arbor where he is currently an Associate Professor in Mechanical Engineering and in Civil and Environmental Engineering. From 2017 to 2018 he was a Visiting Professor in Control and Dynamical Systems at the California Institute of Technology. In 2022 he was a Visiting Professor in Applied Mechanics at the Budapest University of Technology. His research interests include nonlinear dynamics and control, time delay systems, machine learning and data-driven systems with applications to connected and automated vehicles, traffic flow, and biological networks.
\end{IEEEbiography}

% \vspace{-9mm}
\begin{IEEEbiography}
[{\includegraphics[width=1in,height=1.25in,clip,keepaspectratio]{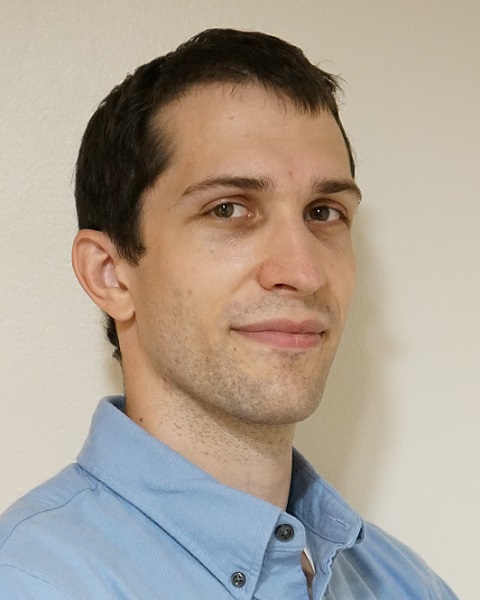}}]
{Tamas G. Molnar} received his BSc degree in Mechatronics Engineering, MSc and PhD degrees in Mechanical Engineering from the Budapest University of Technology and Economics, Hungary, in 2013, 2015 and 2018.
He held postdoctoral position at the University of Michigan, Ann Arbor between 2018 and 2020.
Since 2020 he is a postdoctoral fellow at the California Institute of Technology, Pasadena.
His research interests include nonlinear dynamics and control, safety-critical control, and time delay systems with applications to connected automated vehicles, robotic systems, and machine tool vibrations.
\end{IEEEbiography}

\vfill

\end{document}